\documentclass[a4paper,11pt]{article}
\pdfoutput=1 

\usepackage{dcolumn}
\usepackage{bm}
\usepackage{amsthm}
\usepackage{setspace}
\usepackage[normalem]{ulem}
\usepackage{relsize}
\usepackage{cancel}
\usepackage{verbatim}
\usepackage{enumitem}
\usepackage{mathtools}
\usepackage{empheq}
\usepackage{dsfont}
\usepackage{amsmath}
\usepackage{slashed}
\usepackage{tabu}
\usepackage{graphicx}
\usepackage{caption}
\usepackage{subcaption}

\usepackage{jheppub}

\usepackage{amsthm}

\newcolumntype{I}[1]{>{\centering\arraybackslash$}m{#1}<{$}}

\newcommand{\sdet}{\mathrm{sdet}}

\title{$\mathcal{N}=2$ Liouville SCFT in Four Dimensions}

\author{Tom Levy$^1$,}
\author{Yaron Oz$^1$,}
\author{Avia Raviv-Moshe$^1$}
\affiliation{$^1$ Raymond and Beverly Sackler School of Physics and Astronomy, Tel-Aviv University, 55 Haim Levanon street, Tel-Aviv, 69978, Israel}
\emailAdd{toml@mail.tau.ac.il}
\emailAdd{yaronoz@post.tau.ac.il}
\emailAdd{aviaravi@mail.tau.ac.il}

\abstract{
We construct a Liouville superconformal field theory with eight real supercharges in four dimensions. The Liouville superfield is an $\mathcal{N}=2$ chiral superfield with sixteen bosonic and sixteen fermionic component fields. Its lowest component is a log-correlated complex scalar field whose real part carries a background charge. The theory is non-unitary with a continuous spectrum of scaling dimensions. We study its quantum dynamics on
the supersymmetric 4-sphere and show that the classical background charge is not corrected
quantum mechanically. We calculate the super-Weyl anomaly coefficients and find that $c$ vanishes, while $a$ is negative and depends on the background charge. We derive an integral expression for the correlation functions of superfield vertex operators in $\mathcal{N}=2$ superspace and
analyze them in the semiclassical approximation by using a quaternionic formalism for the $\mathcal{N}=2$ superconformal algebra. }

\begin{document}
\maketitle
\flushbottom

\section{Introduction}

Liouville field theory in two dimensions \cite{Polyakov:1981rd} has been extensively studied for almost three decades and is recognized as  a basic building 
block in quantum field and string theories.
Higher dimensional Liouville field theories appeared recently as a basic ingredient 
of the inertial range field theory of fluid turbulence proposed in \cite{Oz:2017ihc},\cite{Oz:2018mdq},  with the Liouville field being a Nambu-Goldstone boson, the Liouville potential being the local fluid energy dissipation and the Liouville background charge $Q$ related to the fluid intermittency.
The higher even-dimensional  bosonic Liouville field theories were studied  in \cite{Levy:2018bdc} and their  four-dimensional $\mathcal{N}=1$ superconformal
version in \cite{Levy:2018xpu}. 

The aim of this work is to construct  and study $\mathcal{N}=2$ Liouville superconformal field theory (SCFT) in four dimensions. 
The Liouville superfield is an $\mathcal{N}=2$ chiral superfield with sixteen bosonic and sixteen fermionic component fields. Its lowest component is a log-correlated complex scalar field whose real part carries a background charge. The resulting theory is non-unitary with a continuous spectrum of scaling dimensions. 
It localizes semi-classically on solutions that describe curved superspaces with a constant complex supersymmetric $\mathcal{Q}$-curvature. 
We study its quantum dynamics on
the supersymmetric 4-sphere and show that the classical background charge is not corrected
quantum mechanically. This is similar to the non-renormalization of the background charge in the four-dimensional $\mathcal{N}=1$ superconformal
case  \cite{Levy:2018xpu}, and in two-dimensional $\mathcal{N}=2$ Liouville SCFT  \cite{Distler:1989nt,Mussardo:1988av}.

We calculate the super-Weyl anomaly coefficients and find that $c$ vanishes, while $a$ is negative and depends on the background charge. We derive an integral expression for the correlation functions of superfield vertex operators in $\mathcal{N}=2$ superspace and
analyze them in the semiclassical approximation by using a quaternionic formalism for the $\mathcal{N}=2$ superconformal algebra.
In this formalism one can express all superconformal transformations as quaternionic super-M\"obius transformations, i.e. as quaternionic linear fractional transformations. We will use this in order to derive selection rules for the correlation function of vertex operators in the semiclassical limit.

The paper is organized as follows. In section  \ref{sec:Class} we will analyze the classical $\mathcal{N}=2$  Liouville field theory.
 In subsection \ref{subsec:TheModel} we will construct the action of the theory, verify its super-Weyl invariance and
derive the classical field equations. In subsection \ref{subsec:LiouvilleOnSphere} we will study the theory on the  $\mathcal{N}=2$  supermanifold
extension of the 4-sphere where one sees the background charge in the boundary conditions of the Liouville
superfield. We will write the action in components,  check its R-symmetries and construct a solution to the classical
field equations. In section \ref{sec:QuantumSection} we will study the quantum aspects of the theory. We will show that the background charge is not corrected quantum mechanically and we will calculate the super-Weyl anomaly coefficients.  In section \ref{sec:CorrelationFunctionCG} we will study the correlation function of superfield vertex operators and derive an integral expression for them by considering the relation to free fields and
four-dimensional Coulomb gas integrals. In section  \ref{sec:SemiClassicalLimit} we will consider the correlation functions of vertex operators in the semiclassical limit and introduce the quaternionic formalism for $\mathcal{N}=2$ superconformal transformations. We will conclude with a summary and outlook
in section \ref{sec:SummaryAndOutlook}. A summary of notations and details of various calculations are given in the appendices.

\section{$\mathcal{N}=2$ Liouville SCFT  in Four Dimensions}
\label{sec:Class}
In this section we will construct and study the classical aspects of $\mathcal{N}=2$ Liouville superconformal field theory in four dimensions in an $\mathrm{SU}(2)$ superspace. We will mainly use the notations of \cite{Kuzenko:2013gva} and \cite{Kuzenko:2008ep} for the curved supergravity geometrical quantities and the notations of \cite{Butter:2013lta} for flat space notation and reduction to component fields. These are briefly summarized in appendix \ref{app:Notations}. The supergravity notations are in Lorenzian signature $(-,+,+,+)$. We use the analytic continuation in \cite{Festuccia:2011ws} for the Euclidean signature of the supersymmetric 4-sphere.

\subsection{The Classical Theory}
\label{subsec:TheModel}

\subsubsection{The Action}
Four-dimensional $\mathcal{N}=2$ Liouville SCFT is given by the action:
\begin{equation}
\label{eq:ActionMain}
S_L(\Phi, \bar{\Phi}) = \frac{1}{64\pi^2}\int d^4x\, d^4\Theta\, \mathcal{E} \left(\Phi \bar{\Delta}\bar{\Phi}+  2Q\hat{\mathcal{Q}}\Phi + 64\pi^2\mu e^{2b\Phi} \right)+\text{h.c.} \ ,
\end{equation}
where $\mathcal{E}$ is the chiral density \cite{deRoo:1980mm}. The Liouville superfield $\Phi$ is an $\mathcal{N}=2$ chiral superfield \cite{deRoo:1980mm},\cite{deWit:1980lyi}, which satisfies the conditions:
\begin{equation}
\label{eq:ChiralCond}
\bar{\mathcal{D}}^{\dot{\alpha}}_i\Phi =0, \qquad i=\{1,2\} \ , 
\end{equation}
where $\mathcal{D}_A = (\mathcal{D}_a, \mathcal{D}_{\alpha}^{i}, \bar{\mathcal{D}}_{i}^{\dot{\alpha}})$ is the covariant superderivative \cite{Kuzenko:2013gva}. The chiral multiplet $\Phi$ consists of 16 + 16 bosonic and fermionic component fields: a complex scalar field $A$, a chiral spinor doublet $\Psi_i$, a complex symmetric scalar $B_{ij}=B_{ji}$,  an antisymmetric tensor $F_{ab}=-F_{ba}$, a chiral spinor doublet $\Lambda_i$ and a complex scalar field $C$.

The dimensionless parameters in \eqref{eq:ActionMain} are the background charge $Q$, the cosmological constant $\mu$ (which we take to be complex) and $b$. We will denote $S_{C.G.} = \left. S_L \right|_{\mu = 0}$, which describes a free SCFT that we call four-dimensional $\mathcal{N}=2$ Coulomb gas SCFT.

 In \eqref{eq:ActionMain} $\bar{\Delta}$ denotes the chiral projection operator \cite{Kuzenko:2008ry}, \cite{Muller:1989uhj}:
\begin{equation}
\bar{\Delta} = \frac{1}{96}\left( \left(\bar{\mathcal{D}}^{ij}+16\bar{S}^{ij}\right)\bar{\mathcal{D}}_{ij} - \left(\mathcal{\bar{D}}^{\dot{\alpha}\dot{\beta}}-16\bar{Y}^{\dot{\alpha}\dot{\beta}}\right)\bar{\mathcal{D}}_{\dot{\alpha}\dot{\beta}}\right) \ ,
\end{equation}
where $\bar{\mathcal{D}}_{ij} \equiv \bar{\mathcal{D}}_{\dot{\alpha}(i}\bar{\mathcal{D}}^{\dot{\alpha}}_{j)}$ and $\bar{\mathcal{D}}^{\dot{\alpha}\dot{\beta}} = \bar{\mathcal{D}}^{(\dot{\alpha}}_{i}\bar{\mathcal{D}}^{\dot{\beta})i}$. The superfields $S_{ij}=S_{ji}$ and $Y_{\alpha\beta}=Y_{\beta\alpha} $ are complex symmetric torsion tensors (see e.g. \cite{Kuzenko:2013gva} and \cite{Kuzenko:2008ep}). Their complex conjugated partners are denoted by $\bar{S}_{ij}\equiv \overline{S^{ij}}$ and $\bar{Y}^{\dot{\alpha}\dot{\beta}}\equiv \overline{Y^{\alpha\beta}}$. Acting with the chiral projection operator on any scalar superfield $U$ transforms it to a chiral superfield and we have: 
\begin{equation}
\label{eq:ChiralProjOp}
\bar{\mathcal{D}}^{\dot{\alpha}}_{i} \bar{\Delta}U = 0, \quad \int d^4x \,d^4\theta\, d^4\bar{\theta}\, E \, U = \int d^4x\,d^4\Theta\,\mathcal{E}\, \bar{\Delta}U \ ,
\end{equation}
where $E = \sdet\left(E_{A}^{\;\;\;M}\right)$ is the curved superspace integration measure. Using \eqref{eq:ChiralProjOp} we can write the kinetic part of the action \eqref{eq:ActionMain} simply as $\int d^4x \,d^4\theta\, d^4\bar{\theta}\, E \, \Phi\bar{\Phi}$. The chiral projection operator provides an $\mathcal{N}=2$ supersymmetric extension of the conformally covariant fourth-order differential Paneitz operator \cite{Pan}.

In the Liouville action, $\hat{\mathcal{Q}}$ is  an $\mathcal{N}=2$ supersymmetric extension of the conformally covariant $\mathcal{Q}$-curvature  \cite{Q} given by \cite{Butter:2013lta}: 
\begin{equation}
\hat{\mathcal{Q}} \equiv \frac{1}{2}\bar{Y}_{\dot{\alpha}\dot{\beta}}\bar{Y}^{\dot{\alpha}\dot{\beta}}+\frac{1}{2}\bar{S}^{ij}\bar{S}_{ij}+\frac{1}{12}\bar{\mathcal{D}}^{ij}\bar{S}_{ij}.
\end{equation}
This supersymmetric extension is chiral and satisfies $\bar{\mathcal{D}}_{i}^{\dot{\alpha}} \hat{\mathcal{Q}} = 0$.

\subsubsection{Super-Weyl Invariance}

The objects $\bar{\Delta}, \hat{\mathcal{Q}}$ transform covariantly under $\mathcal{N}=2$ super-Weyl transformations \cite{Kuzenko:2008ep}, \cite{Butter:2013lta}, \cite{Kuzenko:2013gva} which are parameterized by a chiral superfield $\sigma$:
\begin{equation}
\label{eq:DeltaQSuperWeyl}
\delta_{\sigma} \bar{\Delta} = -2\sigma \bar{\Delta}, \quad \delta_{\sigma} \hat{\mathcal{Q}} = -2\sigma \hat{\mathcal{Q}} + \bar{\Delta}\bar{\sigma}  \ .
\end{equation}
This transformation law is the super-Weyl generalization of the Weyl transformations of the Paneitz operator and the $\mathcal{Q}$-curvature. Under super-Weyl transformations the Liouville superfield transforms according to
\begin{equation}
\label{eq:PhiSWeylTrans}
\Phi \to \Phi-Q\sigma, \qquad \bar{\Phi} \to \bar{\Phi}-Q\bar{\sigma},
\end{equation}
which, together with \eqref{eq:DeltaQSuperWeyl}, ensures that the action \eqref{eq:ActionMain} is classically invariant under super-Weyl transformation,
\begin{equation}
S_L(\Phi,\bar{\Phi}) \to S_L(\Phi,\bar{\Phi}) - S_{C.G.}(Q\sigma, Q\bar{\sigma}), 
\end{equation}
under the condition that the background charge takes it classical value:
\begin{equation}
\label{eq:QchargeClass}
Q = \frac{1}{b}.
\end{equation}

Similar to the $\mathcal{N}=1$ four dimensional Liouville SCFT studied in \cite{Levy:2018xpu}, the supersymmetric-$\mathcal{Q}$-curvature is related to a topological functional given by \cite{Butter:2013lta}:
\begin{equation}
\label{eq:GuassBonnetFunc}
\int d^4x \,d^4 \Theta \, \mathcal{E} \, \left( 2\hat{\mathcal{Q}}-W^{\alpha\beta}W_{\alpha\beta}\right) = 64\pi^2 (\chi + i p) \ ,
\end{equation}
where the complex symmetric torsion tensor $W_{\alpha\beta} = W_{\beta\alpha}$ is the supersymmetric extension of the Weyl tensor. It transforms homogenously under super-Weyl transformation \cite{Kuzenko:2008ep}:
\begin{equation}
\delta_{\sigma}W_{\alpha\beta} = -\sigma W_{\alpha\beta}. 
\end{equation}
The integral \eqref{eq:GuassBonnetFunc} is a superconformal invariant and when setting to zero the  gravitino and the auxiliary fields of the supergravity multiplet, it is a topological invariant of the resulting curved space, where $\chi$ and $p$ are the
Euler characteristic and first Pontryagin invariant respectively.

In \cite{Kuzenko:2013gva} a Wess-Zumino action for spontaneously broken $\mathcal{N}=2$ superconformal symmetry, whose variation reproduces the $\mathcal{N}=2$ super-Weyl anomaly was introduced. The Goldstone supermultiplet  in \cite{Kuzenko:2013gva} was identified with a reduced chiral superfield \cite{deRoo:1980mm} (i.e. the chiral field strength of an $\mathcal{N}=2$ vector multiplet) containing the dilaton and the axion among its components. This Wess-Zumino action is related to our action \eqref{eq:ActionMain} by replacing the Liouville chiral superfield $\Phi$ with $\frac{1}{b}\log \mathcal{Z}$, where $\mathcal{Z}$ is a reduced chiral superfield satisfying, in addition to \eqref{eq:ChiralCond}, the constraints \cite{deRoo:1980mm} $
\left(\mathcal{D}^{ij}+4S^{ij}\right)\mathcal{Z} = \left(\bar{\mathcal{D}}^{ij}+4\bar{S}^{ij}\right)\bar{\mathcal{Z}} $. There are two significant differences between the two actions: first, the Liouville interaction and the kinetic term 'switch' roles between the two models. This change is technically convenient for the rest of this paper. Second, the number of degrees of freedom carried by the multiplets is different. The reduced chiral superfield carries $8+8$ degrees of freedom (bosonic and fermionic), while the chiral one used in \eqref{eq:ActionMain} carries $16+16$ degrees of freedom. Yet, these two models yield the same super-Weyl variation.

\subsubsection{Field Equations}
The field equations derived from the action \eqref{eq:ActionMain} read:
\begin{equation}
 \bar{\Delta}\bar{\Phi}+Q\hat{\mathcal{Q}}  = -64\pi^2\mu be^{2b\bar{\Phi}} \ , \qquad \Delta\Phi+Q\hat{\bar{\mathcal{Q}}}=-64\pi^2\bar{\mu}be^{2b\Phi}.
\end{equation}
Using the finite form of the transformation \eqref{eq:DeltaQSuperWeyl} (see \cite{Kuzenko:2008qw} for the finite super-Weyl transformation of the relevant curved superspace geometrical quantities) one sees that solutions to the field equations describe super-Weyl parameters $\sigma = b\Phi$ and $\bar{\sigma}=b\bar{\Phi}$  that transform the background to a supermanifold with a constant complex super-$\mathcal{Q}$-curvature:
\begin{equation}
\hat{\mathcal{Q}}  = -64\pi^2\mu b^2, \qquad \bar{\hat{\mathcal{Q}}}=-64\pi^2\bar{\mu} b^2.
\end{equation}
This result is similar to the one found in \cite{Levy:2018xpu} for the $\mathcal{N}=1$ case.  Note, however, that the field equations in the non-supersymmetric four-dimensional Liouville field theory studied in \cite{Levy:2018bdc} have a real positive cosmological constant parameter $\mu$ and their solutions can
be viewed as Weyl factors that transform the background curved space into a constant negative $\mathcal{Q}$-curvature one.

\subsection{Liouville $\mathcal{N}=2$ SCFT on $S^4$}
\label{subsec:LiouvilleOnSphere}
\subsubsection{Background Charge }

We define Liouville SCFT on the $\mathcal{N}=2$ supermanifold extension of $S^4$ \cite{Butter:2015tra}. Using the fact that this supermanifold is superconformally flat, i.e. $W_{\alpha\beta} =0$, and the topological invariant \eqref{eq:GuassBonnetFunc}, where for supersymmetric $S^4$ we have $\chi=2, p=0$, we get that for a constant shift by $\phi_0$ of the Liouville superfield:
\begin{equation}
\label{eq:ActionShiftBC}
S_L(\Phi+\phi_0,\bar{\Phi}+\bar{\phi_0}) = S_L(\Phi,\bar{\Phi}) + 4Q\mathrm{Re}(\phi_0) \ .
\end{equation}
Preforming a singular super-Weyl transformation from supersymmetric $S^4$ to flat superspace and using the transformation law \eqref{eq:PhiSWeylTrans} we find the following boundary conditions for the Liouville superfield:
\begin{equation}
\label{eq:Boundary1Phi}
\Phi = -2Q\log(|x|) + O(1), \quad |x|\to \infty \ .
\end{equation}
Writing \eqref{eq:Boundary1Phi} in component fields, one finds:
\begin{equation}
\label{eq:BoundaryA1}
\mathrm{Re}(A)=-2Q\log(|x|)+O(1), \quad |x| \to \infty,
\end{equation}
while all other component fields approach a finite limit. Following \eqref{eq:BoundaryA1}, the real part of the lowest component $A$ of the Liouville superfield $\Phi$, i.e. $\mathrm{Re}(A)$, acquires a background charge $Q$.

\subsubsection{Action in Components, R-Symmetry and a Classical Solution}

The kinetic part of the Lagrangian in flat space written in terms of component fields is: 
\begin{equation}
\begin{aligned}
\label{eq:KineticComponents}
&\frac{1}{32\pi^2} \int d^4\theta d^4\bar{\theta} \Phi\bar{\Phi} = \\
 &\frac{1}{32\pi^2}\left( 4\square A \square A^\dagger + CC^\dagger-\bar{B}_{ij}\square B^{ij} -8\partial_aF^{-ab}\partial^cF^+_{cb}+2\bar{\Psi}_i\square\slashed{\partial}\Psi^i -2\bar{\slashed{\partial}\Lambda_i}\Lambda^i\right) .
 \end{aligned} 
\end{equation}
Its bosonic parts can be found in \cite{deWit:2010za}. The interaction term in \eqref{eq:ActionMain} can be derived in terms of component fields by constructing the chiral superfield $e^{\Phi}$ in terms of the components of $\Phi$. For simplicity and for later use, we list here only the bosonic parts of the interaction. 
The bosonic parts of the Liouville interaction yield:
\begin{equation}
\begin{aligned}
& \mu\int d^4xd^4\theta\mathcal{E}e^{2b\Phi}+h.c.=\\
&-\mu b\left(e^{2bA}C-\frac{b}{2}e^{2bA}B_{ij}B^{ij}+bF^{-ab}F^-_{ab}e^{4bA}\right)+h.c.
\end{aligned}
\end{equation}
Four-dimensional Liouville $\mathcal{N}=2$ SCFT is invariant under the $\mathrm{SU}(2)_R\times \mathrm{U}(1)_R$ R-symmetry group. As in the $\mathcal{N}=1$ case 
\cite{Levy:2018bdc},  the $\mathrm{U}(1)_R$ symmetry is not a standard one (the interaction term, as well as the boundary condition \eqref{eq:Boundary1Phi} are not invariant under a standard $\mathrm{U}(1)$ transformation). Rather, the Liouville superfield transforms in the affine representation under a non-compact $\mathrm{U}(1)_R$ symmetry:
\begin{equation}
\label{eq:SuperspaceRSymm}
\Phi'\left(e^{i\alpha/2}\theta,x_+\right) = \Phi(\theta,x_+)+\frac{i}{b}\alpha, \quad \; \;
\bar{\Phi}'\left(e^{-i\alpha/2}\bar{\theta},x_-\right) = \bar{\Phi}\left(\bar{\theta},x_-\right)-\frac{i}{b}\alpha \ .
 \end{equation}
In terms of bosonic component fields, the transformation \eqref{eq:SuperspaceRSymm} reads:
\begin{equation}
A \to A +\frac{i}{b}\alpha,\quad B_{ij}\to e^{-i\alpha}B_{ij} \quad F^-_{ab} \to e^{-i\alpha}F^-_{ab},\quad C \to e^{-2i\alpha}C. 
\end{equation}
Under the $\mathrm{SU}(2)_R$ symmetry the Liouville chiral superfield transforms trivially. The fermions $\Psi^i$ and $\Lambda^i$ transform as doublets under the $\mathrm{SU}(2)_R$ transformation, while the symmetric complex field $B^{ij}$ transforms in the triplet representation under $\mathrm{SU}(2)_R$. All other component fields transform as singlets under $\mathrm{SU}(2)_R$.

The classical equations of motion with vanishing fermions are:
\begin{equation}
\label{eq:FieldEOM}
\square^2A+8\pi^2\bar{\mu}b^3e^{2b\bar{A}}\bar{B}_{ij}\bar{B}^{ij}=0,
\end{equation}
\begin{equation}
-\square B_{ij}+32\pi^2\bar{\mu}b^2 e^{2b\bar{A}}\bar{B}_{ij}=0,
\end{equation}
\begin{equation}
C=32\pi^2\bar{\mu} b e^{2b\bar{A}},
\end{equation}
\begin{equation}
\partial^c\partial_a F^{-ab}+8\pi^2\bar{\mu}bF^{+cb}e^{4b\bar{A}}=0.
\end{equation}
Therefore, a possible solution to the classical equations of motion  that respects the boundary condition \eqref{eq:BoundaryA1} is given by:
\begin{equation}
\label{eq:SolToClassicalEOMRegularNotation}
\begin{aligned}
&A = -\frac{1}{b}\log{\left({4\pi^2|\mu|b^2}|x|^2+1\right)}, \qquad C= \frac{32\pi^2\bar{\mu} b}{\left({4\pi^2|\mu|b^2}|x|^2+1\right)^2},\\
&B_{ij} = \frac{4\pi i\sqrt{6\bar{\mu}}}{\left(4\pi^2{|\mu|b^2}|x|^2+1\right)}\delta_{ij}, \qquad \bar{B}_{ij} = -\frac{4\pi i\sqrt{6\mu}}{\left(4\pi^2{|\mu|b^2}|x|^2+1\right)}\delta_{ij}.
\end{aligned}
\end{equation}
All other fields vanish.

\section{Quantum $\mathcal{N}=2$ Liouville SCFT in Four Dimensions}
\label{sec:QuantumSection}
In this section we study quantum aspects of the Liouville $\mathcal{N}=2$ SCFT in four dimensions. We show that the background charge is not corrected quantum mechanically and calculate supersymmetric Weyl anomaly coefficients. We find that the $c$ anomaly coefficient vanishes while the $a$ anomaly coefficient is negative and depends on the value of the background charge $Q$.

\subsection{Primary Vertex Operators and Background Charge Non-Renormalization}
The free field two-point function of the lowest order component field reads:
\begin{equation}
\left\langle A(x)\bar{A}(x') \right\rangle_{C.G.}=-\log(|x-x'|) \ ,
\end{equation}
thus, it is a log-correlated complex field.
Consider the vertex operator:
\begin{equation}
\label{eq:BosonicVertexOp}
V_{\alpha\tilde{\alpha}} = e^{2\alpha A+2\tilde{\alpha}\bar{A}} = e^{2(\alpha+\tilde{\alpha})\mathrm{Re}(A) + 2i(\alpha-\tilde{\alpha})\mathrm{Im}(A)},
\end{equation}
where $\alpha, \tilde{\alpha}$ are two independent complex numbers. When $\alpha = \tilde{\alpha}$ we denote $V_{\alpha}=V_{\alpha\tilde{\alpha}}$. The vertex operators have scaling dimensions:
\begin{equation}
\label{eq:DimBosonicVertexOp}
\Delta_{\alpha\bar{\alpha}} = -4\alpha\tilde{\alpha}+2Q(\alpha+\tilde{\alpha}) \ .
\end{equation}
Requiring the Liouville interaction to be a marginal operator implies:
\begin{equation}
\Delta(e^{2bA}) = \Delta_{b,0} = 2 \ ,
\end{equation}
and using \eqref{eq:DimBosonicVertexOp} we get the quantum value of the background charge:
\begin{equation}
\label{eq:NonRenormQ}
Q = \frac{1}{b} \ .
\end{equation}
We see that the classical value of the background charge \eqref{eq:QchargeClass} is not corrected quantum mechanically.
Note that  $\mathcal{N}=1$ Liouville SCFT in four dimensions \cite{Levy:2018xpu}
and $\mathcal{N}=2$ Liouville SCFT in two dimensions \cite{Distler:1989nt,Mussardo:1988av} exhibit a similar non-renormalization  of the background charge. 

We can also consider vertex operators in superspace:
\begin{equation}
\label{eq:GeneralVertexInSuperspace}
\mathcal{V}_{\alpha\tilde{\alpha}} = e^{2\alpha \Phi +2\tilde{\alpha}\bar{\Phi}}.
\end{equation}
Its dimension is given by \eqref{eq:DimBosonicVertexOp} and its $U(1)_R$ charge reads:
\begin{equation}
\label{eq:U1RGenVertexInSuperspace}
w_{\alpha\tilde{\alpha}} = \frac{2}{b}(\alpha-\tilde{\alpha}).
\end{equation}
The following operator is a chiral primary operator:
\begin{equation}
\label{eq:ChiralVertexOp}
\mathcal{V}_\alpha \equiv \mathcal{V}_{\alpha,0}= e^{2\alpha\Phi}.
\end{equation}
Its dimension is equal to its $U(1)_R$ charge and they are given by $\Delta_{\mathcal{V}_\alpha} = w_{_{\mathcal{V}_{\alpha}}} = \frac{2}{b}\alpha$.

The free-fields superspace propagators are given by:
\begin{equation}
\label{eq:PropSuper01}
\left<\Phi(x,\theta^i,\bar{\theta}_j)\Phi(x',\theta^{'i},\bar{\theta}^{'}_j)\right> = 0 \ ,
\end{equation}
\begin{equation}
\label{eq:PropSuper02}
\left<\bar{\Phi}(x,\theta^i,\bar{\theta}_j)\bar{\Phi}(x',\theta^{'i},\bar{\theta}^{'}_j)\right> = 0 \ ,
\end{equation}
\begin{equation}
\label{eq:PropSuperfields}
\left<\Phi(x,\theta^i,\bar{\theta}_j)\bar{\Phi}(x',\theta^{'i},\bar{\theta}^{'}_j)\right> = e^{\left(\theta\sigma^a\bar{\theta}+\theta'\sigma^a\bar{\theta}'-2\theta\sigma^a\bar{\theta}'\right)\partial_a^x}\left\langle A(x)\bar{A}(x') \right\rangle,
\end{equation}
where $\theta\sigma^a\bar{\theta} = \theta^i\sigma^a\bar{\theta}_i= \theta^1\sigma^a\bar{\theta}_1+\theta^2\sigma^a\bar{\theta}_2$.
For a detailed derivation of \eqref{eq:PropSuper01}-\eqref{eq:PropSuperfields} see appendix \ref{app:SuperSpacePropagators}. For future reference, in supersymmetric Coulomb gas theory the following identity holds:
\begin{equation} \label{eq:SuVerOpWickCon}
 \left<\mathcal{V}_{\alpha_1\tilde{\alpha}_1}(z_1)\cdots \mathcal{V}_{\alpha_N\tilde{\alpha}_N}(z_N)\right>_{C.G.} = \prod_{l\neq m}|z_l-\bar{z}_m|^{-4\alpha_l\tilde{\alpha}_m} \ , 
 \end{equation} 
where $|z_l - \bar{z}_m|$ is the distance in superspace between a chiral and an anti-chiral coordinate:
\begin{equation}
|z_l -\bar{z}_m|^2 = \left(x_l-x_m +i\theta_l\sigma\bar{\theta}_l + i\theta_m\sigma\bar{\theta}_m -2\theta_l\sigma\bar{\theta}_m\right)^2 =\left( x_{l+}-x_{m-} - 2\theta_l\sigma\bar{\theta}_m\right)^2 \ .
\end{equation}
This correlation function vanishes unless $\sum\alpha_l = \sum\tilde{\alpha}_l = \frac{1}{b}$.

\subsection{$\mathcal{N}=2$ Super-Weyl Anomalies}

The super-Weyl anomaly can be related to the ordinary Weyl anomaly by
setting the gravitino and the auxiliary fields of the supergravity multiplet to zero. The anomaly coefficients $a$ and $c$ correspond to the A-type and B-type Weyl anomalies \cite{Deser:1993yx}.

The Weyl anomaly coefficients of the Liouville SCFT do not depend on the interaction terms and are the same as those of the supersymmetric Coulomb gas theory. The latter is a free SCFT, which consists of an ordinary four-dimensional Coulomb gas CFT $\mathrm{Re}(A)$ with a background charge $Q$, a conformal four-derivative real scalar $\mathrm{Im}(A)$,  a conformal three-derivative chiral spinor doublet $\Psi_i$, a conformal two-derivative complex symmetric scalar $B_{ij}$, a conformal antisymmetric tensor $F_{ab}$, a conformal one-derivative chiral spinor doublet $\Lambda_i$ and an auxiliary complex scalar field $C$.  Thus, to obtain the super-Weyl anomaly of Liouville SCFT, we should sum the anomalies of the component fields:

\bgroup
\def\arraystretch{1.3}
\begin{center}
  \begin{tabular}{| c | c | c | }
    \hline
    Field & $a$ & $c$ \\
    \hline
    $\mathrm{Re}(A)$ & $ -\frac{7}{90}-Q^2$ & $-\frac{1}{15}$ \\ \hline
    $\mathrm{Im}(A)$ & $-\frac{7}{90}$ & $-\frac{1}{15}$ \\
     \hline
    $\Psi_i $ & $-\frac{3}{40}$ & $-\frac{1}{60}$ \\  
      \hline
    $ B_{ij}$ & $\frac{1}{60}$ & $\frac{1}{20}$ \\ 
    \hline 
    $F_{ab}$ & $-\frac{19}{60}$ & $\frac{1}{20}$ \\
    \hline
    $\Lambda_i$ & $\frac{11}{360}$ & $\frac{1}{20}$ \\
    \hline
    $C$ & $0$ & $0$ \\
    \hline 
  \end{tabular}
\end{center}
\egroup

Summing up the coefficients we obtain:
\begin{equation}
a = -\frac{1}{2}-Q^2, \quad c = 0 \ .
\end{equation}

\section{Correlation Functions and Supersymmetric Coulomb Gas Integrals}
\label{sec:CorrelationFunctionCG}

In this section we study the correlation functions of superfield vertex operators \eqref{eq:ChiralVertexOp} in four-dimensional $\mathcal{N}=2$  Liouville SCFT by using their relation to the free supersymmetric Coulomb gas theory and derive an integral expression for them. We consider general correlation functions of these operators:
\begin{equation}
\label{eq:DefCorrMain}
\mathcal{G}_{\alpha_1\tilde{\alpha}_1,\dots,\alpha_N\tilde{\alpha}_N} (z_1,\dots,z_N) = \left<\mathcal{V}_{\alpha_1\tilde{\alpha}_1}(z_1)\cdots \mathcal{V}_{\alpha_N\tilde{\alpha}_N}(z_N)\right>,
\end{equation}
where the expectation value is defined as:
\begin{equation} \label{eq:VertexCorrelationFunc}
\left<\mathcal{V}_{\alpha_1\tilde{\alpha}_1}(z_1)\cdots \mathcal{V}_{\alpha_N\tilde{\alpha}_N}(z_N)\right> \equiv \int D\Phi\, e^{-S_L}\prod_{l=1}^{N} e^{2\alpha_{l}\Phi(z_l)+2\tilde{\alpha}_l\bar{\Phi}(z_l)} \ .
\end{equation}
We consider the shift $A \to A-\frac{\log \mu}{2b}$ and using \eqref{eq:ActionShiftBC} we obtain a KPZ scaling relation:
\begin{equation} \label{eq:KPZScale}
\mathcal{G}_{\alpha_1\tilde{\alpha}_1,\dots,\alpha_N\tilde{\alpha}_N} (x_1,\dots,x_N) \propto \mu^{s}\bar{\mu}^{\tilde{s}},~~~~s = \frac{1/b-\sum_{l} \alpha_l}{b},~~~~\tilde{s}= \frac{1/b-\sum_{l} \tilde{\alpha}_l}{b} \ .
\end{equation}
As a result of the KPZ scaling relation \eqref{eq:KPZScale} one can see that the correlation functions are not analytic functions of the cosmological constant $\mu,\bar{\mu}$. This implies that we can not preform a naive perturbation theory calculation in powers of $\mu$. We write the interaction terms of the action \eqref{eq:ActionMain} using the chiral and anti-chiral integrals:
\begin{equation}
\label{eq:ChiralActions}
S_+ = \int d^4x\, d^4\theta\, e^{2b\Phi}, \quad
S_- = \int d^4x\, d^4\bar{\theta}\, e^{2b\bar{\Phi}}  \ ,
\end{equation}
so that $S_L = S_{C.G.}+\mu S_++\bar{\mu}S_-$. We can separate the real zero mode $\Phi_0\in\mathbb{R}$ of the path integral over the real part of $A(x)$ from the non-zero mode $\widehat{A}(x)$ and write the corresponding superfield decomposition $\Phi(z)=\Phi_0+\widehat{\Phi}(z)$
(see the two-dimensional bosonic case in e.g. \cite{Teschner:2001rv}) . Integrating over the zero mode using equation \eqref{eq:ActionShiftBC} we get the following expression for the correlation functions:
\begin{equation}
\label{eq:CorrFuncGammaInt}
\begin{aligned}
&\mathcal{G}_{\alpha_1\tilde{\alpha}_1,\dots,\alpha_N\tilde{\alpha}_N} (x_1,\dots,x_N) = \\
&= \int d\Phi_0\, D\widehat{\Phi}\, e^{-S_L\left(\Phi_0+\widehat{\Phi}\right) }\prod_{l=1}^{N} e^{2\alpha_{l}\left(\Phi_0+\widehat{\Phi}(z_l)\right)+2\tilde{\alpha}_l\left(\Phi_0+\widehat{\bar{\Phi}}(z_l)\right)}\\
&=  \frac{\Gamma(-s-\tilde{s})}{2b}\left<\prod_{l=1}^{N}e^{2\alpha_{l}\widehat{\Phi}(z_l)+2\tilde{\alpha}_l\widehat{\bar{\Phi}}(z_l)}(\mu S_++\bar{\mu} S_-)^{s+\tilde{s}}\right>_{C.G.} \ ,
\end{aligned}
\end{equation}
where we denote by $\left<\dots\right>_{C.G.}$ expectation values in the free Coulomb gas SCFT. The gamma function appearing in \eqref{eq:CorrFuncGammaInt} produces poles in the correlation functions \eqref{eq:DefCorrMain} when  $s+\tilde{s}$ is a non-negative integer, i.e. $s+\tilde{s} = k\in\mathbb{N}\cup\{0\}$. As a consequence of $\eqref{eq:CorrFuncGammaInt}$, the residues of the correlation function at these poles, when considered as a function of the variable $2\sum(\alpha_l+\tilde{\alpha}_l)$, are given exactly by the the result of the naive perturbation theory in $\mu, \bar{\mu}$:
\begin{equation}
\mathcal{G}^{(k)}_{\alpha_1\tilde{\alpha}_1,\dots,\alpha_N\tilde{\alpha}_N} (z_1,\dots,z_N) = \;\underset{2\sum(\alpha_l+\tilde{\alpha}_l)=\frac{4}{b}-2kb}{\mathrm{Res}}\;\mathcal{G}_{\alpha_1\tilde{\alpha}_1,\dots,\alpha_N\tilde{\alpha}_N} (z_1,\dots,z_N) \ ,
\end{equation}  
where the LHS denotes the contribution of $k$th-order naive perturbation theory to the correlation functions: 
\begin{equation}
\label{eq:GResidueDecomp}
\begin{aligned}
\mathcal{G}^{(k)}_{\alpha_1\tilde{\alpha}_1,\dots,\alpha_N\tilde{\alpha}_N} (z_1,\dots,z_N)
&=\frac{1}{k!}\left<\prod_{l=1}^{N}\mathcal{V}_{\alpha_l\tilde{\alpha}_l}(z_l)(-\mu S_+-\bar{\mu} S_-)^k\right>_{C.G.} \\
&=\sum_{n+\tilde{n} = k} \frac{(-\mu)^n(-\bar{\mu})^{\bar{n}}}{n!\tilde{n}!}\left<\prod_{l=1}^{N}\mathcal{V}_{\alpha_l\tilde{\alpha}_l}(z_l)(S_+)^{n}(S_-)^{\tilde{n}}\right>_{C.G.}\\
&\equiv \sum_{n+\tilde{n}=k}(-\mu)^n(-\bar{\mu})^{\tilde{n}}\mathcal{G}^{(n,\tilde{n})}_{\alpha_1\tilde{\alpha}_1,\dots,\alpha_N\tilde{\alpha}_N} (z_1,\dots,z_N) .
\end{aligned}
\end{equation}
The KPZ scaling relation \eqref{eq:KPZScale} shows that actually only a single term in the sum \eqref{eq:GResidueDecomp} is non-vanishing. Therefore, the correlation functions have a pole only when both $s$ and $\tilde{s}$ are non-negative integers, which we denote by $s=n,\;\tilde{s}=\tilde{n}$.

 We now focus on the three-point function of vertex operators, i.e. $N=3$.  We can write explicit integral expressions for the residues \eqref{eq:GResidueDecomp} corresponding to the three-point function by using the free field correlation functions  \eqref{eq:SuVerOpWickCon} to write:
\begin{equation}
\label{eq:GResiduePsiFDecomp}
\begin{aligned}
&\mathcal{G}^{(n,\tilde{n})}_{\alpha_1\tilde{\alpha}_1,\alpha_2\tilde{\alpha}_2,\alpha_3\tilde{\alpha}_3} (z_1,z_2,z_3) = \frac{1}{n!\tilde{n}!} \prod_{l\neq m}|z_l-\bar{z}_m|^{-4\alpha_l\tilde{\alpha}_m} \int d^{8n}z'\,d^{8\tilde{n}}\bar{z}'  \\
& \times \prod_{p,\tilde{p}}|z'_p-\bar{z}'_{\tilde{p}}|^{-4b^2}\prod_{l=1}^{3}\prod_{\tilde{p}=1}^{\tilde{n}}|z_l-\bar{z}'_{\tilde{p}}|^{-4b\alpha_l}\prod_{p=1}^{n}|z'_p-\bar{z}_l|^{-4b\tilde{\alpha}_l} \ ,
\end{aligned}
\end{equation}
where the measures are $d^{8n}z' = \prod_{p=1}^n d^{4}x'_{p+}d^{4}\theta_p'$, $\;d^{8\tilde{n}}\bar{z}' = \prod_{\tilde{p}=1}^n d^{4}x_{\tilde{p}-}'d^{4}\bar{\theta}_{\tilde{p}}'$. One case for which the integral can be explicitly computed is for the first pole $n=0, \tilde{n} = 1$ of the three-point function. According to \eqref{eq:KPZScale}, at the pole $n=0$ we have the relation $\sum_l 2b\alpha_l =2$. The integral can then be written as:
\begin{equation}
\mathcal{G}^{(0,1)}_{\alpha_1\tilde{\alpha}_1,\alpha_2\tilde{\alpha}_2,\alpha_3\tilde{\alpha}_3} (z_1,z_2,z_3) = \prod_{l\neq m}|z_l-\bar{z}_m|^{-4\alpha_l\tilde{\alpha}_m}\int d^4x_- d^4\bar{\theta} \prod_{l=1}^{3} |z_l-\bar{z}|^{-4b\alpha_l} \ .
\end{equation}
This integral is an $\mathcal{N}=2$ generalization of the conformal and $\mathcal{N}=1$ superconformal integrals considered in \cite{Osborn:1998qu}. We define $z_l^2 \equiv |z_l-\bar{z}|^2 = (x_{l+}-x_{-}-2i\theta_l^i\sigma\bar{\theta}_{li})^2$ and the integral we need to calculate is:
\begin{equation}
I_{\alpha_1\alpha_2\alpha_3} = \int d^4x_{-} d^4\bar{\theta} \prod_{l=1}^{3} \frac{1}{(z_l^2)^{2b\alpha_l}} \ .
\end{equation}
Using the identity $(x^2+i\epsilon)^{-\alpha} =\frac{e^{i\pi\alpha/2}}{\Gamma(\alpha)} \int_0^{\infty}d\lambda\, \lambda^{\alpha-1} e^{i\lambda x^2}$ we have:
\begin{equation}
\begin{aligned}
 I_{\alpha_1\alpha_2\alpha_3}&=\frac{1}{\prod_{l=1}^{3}\Gamma(2b\alpha_l)}\int_0^{\infty} \prod_{l=1}^{3}d\lambda_l\, \lambda_l^{2b\alpha_l-1}\int d^4x_{-}d^4\bar{\theta}\,e^{i\sum_{l=1}^{3}\lambda_l z_l^2}  \\
 &= \frac{\pi^2}{\prod_{l=1}^{3}\Gamma(2b\alpha_l)}\int_0^{\infty} \prod_{l=1}^{3}d\lambda_l\, \lambda_l^{2b\alpha_l-1}\int d^4\bar{\theta}\,\exp\left(-\frac{1}{2\Lambda}\sum_{l,m}\lambda_l\lambda_m z_{lm}^2\right) \ , \\
 \end{aligned}
\end{equation}
where we have defined  $z_{lm} = x_{l+}-x_{m+} - 2i(\theta_l^i-\theta_m^i)\sigma\bar{\theta}_{i}$, $\Lambda = \sum_l \lambda_l$ and preformed the Gaussian integration over $x_-$. Preforming the integration over $\bar{\theta}$ we have:
\begin{equation}
\begin{aligned}
 I_{\alpha_1\alpha_2\alpha_3} &= \frac{\pi^2}{\prod_{l=1}^{3}\Gamma(2b\alpha_l)}\int_0^{\infty} \prod_{l=1}^{3}d\lambda_l\, \lambda_l^{2b\alpha_l-1}\frac{1}{\Lambda^2}\cdot\\
 & \sum_{l,m,p,q}\theta^1_l\sigma\cdot\partial_l \bar{\sigma}\cdot\partial_m\theta^1_m\theta^2_p\sigma\cdot\partial_p \bar{\sigma}\cdot\partial_q\theta^2_q e^{-\frac{1}{2\Lambda}\sum_{l,m}\lambda_l\lambda_m z_{lm}^2} \ . 
  \end{aligned}
\end{equation}

The result of this integral is proportional to the pole of the correlation function of vertex operators \eqref{eq:VertexCorrelationFunc}. As such, the integral should be proportional to an $\mathcal{N}=2$ chiral superconformal invariant built out of the supercoordinates, which we denote $\mathcal{I}_{\alpha_1\alpha_2\alpha_3}(z_1,z_2,z_3)$. We are interested in the dependence of the proportionality constant $C(\alpha_1,\alpha_2,\alpha_3)$ on $\alpha_l$. To obtain the dependence, we use the result \cite{Symanzik:1972wj}:
\begin{equation}
\int_0^{\infty} \prod_{l=1}^{3}d\lambda_l\, \lambda_l^{\delta_l-1}\frac{1}{\Lambda^P} e^{-\frac{1}{2\Lambda}\sum_{l,m}\lambda_l\lambda_m u_{lm}^2} = \frac{\prod_{l=1}^{3}\Gamma(P-\delta_l)}{u_{12}^{P-\delta_3}u_{13}^{P-\delta_2}u_{23}^{P-\delta_1}} \ ,
\end{equation}
for $u_{lm}=u_{ml}$ and $\sum\delta_l = 2P$. Applying this result for our integral we get the proportionality constant:
\begin{equation}
C(\alpha_1,\alpha_2,\alpha_3) = \pi^2\frac{\prod_{l=1}^{3}\Gamma(4-2b\alpha_l)}{\prod_{l=1}^{3}\Gamma(2b\alpha_l)} \ ,
\end{equation}
which is a particular example of the DOZZ three-point function formula \cite{Zamolodchikov:1995aa},\cite{Dorn:1992at}.
and its four-dimensional analogue \cite{Levy:2018bdc},\cite{Furlan:2018jlv}.  
We can thus write the first pole of the correlation function as:
\begin{equation}
\mathcal{G}^{(0,1)}_{\alpha_1\tilde{\alpha}_1,\alpha_2\tilde{\alpha}_2,\alpha_3\tilde{\alpha}_3} (z_1,z_2,z_3) = \pi^2\frac{\prod_{l=1}^{3}\Gamma(4-2b\alpha_m)}{\prod_{l=1}^{3}\Gamma(2b\alpha_m)}\prod_{l\neq m}|z_l-\bar{z}_m|^{-4\alpha_l\tilde{\alpha}_m}\mathcal{I}_{\alpha_1\alpha_2\alpha_3}(z_1,z_2,z_3) \ .
\end{equation}

\section{Correlation Functions in the Semiclassical Limit}
\label{sec:SemiClassicalLimit}
 In this section we consider the correlation functions of vertex operators \eqref{eq:DefCorrMain} in the semiclassical limit $b\to 0$. In this limit we define the rescaled Liouville superfield $\Phi_c = b\Phi$, in terms of which the correlations function that we wish to evaluate are:
 \begin{equation}
 \label{eq:SemiclassCorrFunc}
 \left<\mathcal{V}_{\alpha_1\tilde{\alpha}_1}(z_1)\cdots \mathcal{V}_{\alpha_N\tilde{\alpha}_N}(z_N)\right> \equiv \int D\Phi_c\, e^{-S_L}\prod_{l=1}^{N} \exp\left({2\frac{\alpha_{l}}{b}\Phi_c(z_l)+2\frac{\tilde{\alpha}_l}{b}\bar{\Phi}_c(z_l)}\right) \ .
 \end{equation}
Noting that written in terms of $\Phi_c$ the action \eqref{eq:ActionMain} scales as $S_L\sim b^{-2}$, we can use the saddle point approximation to evaluate the integral \eqref{eq:SemiclassCorrFunc} in the semiclassical limit $b\to 0$. If the inserted vertex operators $\mathcal{V}_{\alpha\tilde{\alpha}}$ obey the scaling $\alpha,\tilde{\alpha}\sim b$ then the insertions do not affect the saddle point of the integral, which is then determined solely by minimizing the action. Such operators are called light operators, and for those we write $\alpha = b\sigma,\; \tilde{\alpha}=b\tilde{\sigma}$, where $\sigma, \tilde{\sigma}$ are kept fixed in the limit $b\to 0$  \cite{Zamolodchikov:1995aa},\cite{Harlow:2011ny}.

Considering the correlation functions of light vertex operators, the  leading exponential asymptotic behaviour in the limit $b\to 0$ is given by the semiclassical expression:
\begin{equation}
\label{eq:SemiclassCorrExp}
\left<\mathcal{V}_{b\sigma_1 b\tilde{\sigma}_1}(z_1)\cdots \mathcal{V}_{b\sigma_N b\tilde{\sigma}_N}(z_N)\right> \sim  e^{-S_L(\Phi_c,\Phi_c)} \int_{\mathcal{M}} d\mu(g)\,\prod_{l=1}^{N} e^{2\sigma_l\Phi_{g}(z_l)+2\tilde{\sigma}_l\bar{\Phi}_{g}(z_l)} \ ,
\end{equation}
where we have assumed that there is a continuum of saddle points and therefore the saddle point approximation must include an integral over them. In \eqref{eq:SemiclassCorrExp}, $\mathcal{M}$ is the moduli space of solutions to the field equations \eqref{eq:FieldEOM} equipped with some coordinates $\lbrace g \rbrace$, $d\mu(g)$ is a measure over this space and $\Phi_g(z), \bar{\Phi}_g(z)$ are the Liouville superfield saddle points as functions of the moduli space coordinates. In addition, $S_L(\Phi_c,\Phi_c)$ is the minimal value of the action, which is the value of the action evaluated at any saddle point.

Using the four-dimensional $\mathcal{N}=2$ superconformal invariance of the theory, given a solution to the field equations \eqref{eq:SolToClassicalEOMRegularNotation}, one can produce further solutions by applying superconformal transformations to the original solution. The moduli space of solutions to the field equations is given by the orbit of the solution  \eqref{eq:SolToClassicalEOMRegularNotation} under the action of the  $\mathcal{N}=2$ superconformal group. We can analyse and integrate over this moduli space by using the four-dimensional $\mathcal{N}=2$ super-M\"{o}bius representation of the superconformal transformations.

\subsection{Four-Dimensional $\mathcal{N}=2$ Super-M\"{o}bius Transformations}
\label{subsec:ForDNequals2TransQuanternions}

In \cite{Lukierski:1983jg} it was shown that the $\mathcal{N}=2$ superconformal group in Euclidean four dimensions can be identified with a quaternionic supergroup $\mathrm{SL}(2|1;\mathbb{H})$. This supergroup has the quaternionic supermatrix representation:
\begin{equation}
\label{eq:SLquaternion}
\mathrm{SL}(2|1;\mathbb{H}) = \left\lbrace \left. M = \begin{pmatrix} 
a & b & \gamma \\
c & d & \delta \\
\alpha & \beta & e
\end{pmatrix} \; \right| \; a,b,c,d,e \in \mathbb{H},\;\, \alpha,\beta,\gamma, \delta \in \mathbb{H}_a, \;\, \sdet(M) =1 \right\rbrace \ ,
\end{equation}
where $\mathbb{H}$ denotes standard quaternions and $\mathbb{H}_a$ denotes Grassmann quaternions, i.e. quaternions whose coefficients of the basis elements are anticommuting (Grassmann) numbers. The quaternionic superdeterminant, $\sdet(M)$, is defined by first identifying each quaternion or Grassmann quaternion with a $2\times 2$ complex or Grassmann complex matrix respectively,
\begin{equation}
q = q_0 + q_1 i +q_2 j + q_3 k \cong \begin{pmatrix}
q_0 + q_1 i & \; q_2 + q_3 i \\
-q_2 +q_3 i & \; q_0 - q_1 i \\
\end{pmatrix} \ , 
\end{equation}
which results in a $(4|2)\times(4|2)$ complex supermatrix, and then taking the standard superdeterminant of the resulting supermatrix.

The quaternionic Lie superalgebra $\mathfrak{sl}(2|1,\mathbb{H})$ is isomorphic to the four-dimensional   $\mathcal{N}=2$ Euclidean superconformal algebra. This was proved in \cite{Lukierski:1983jg} by explicitly writing the quaternionic supermatrix generators of $\mathrm{SL}(2|1;\mathbb{H})$, computing their superalgebra and showing its equivalence to the $\mathcal{N}=2$ superconformal algebra (the calculation is summarized  in appendix \ref{app:QuaternionicAlgebra}).

Based on the isomorphism between the two supergroups, we are able to express all superconformal transformations as quaternionic super-M\"{o}bius transformations, i.e. as quaternionic linear fractional transformations. This provides a realization of the isomorphism on transformations of the supercoordinates. To do so, we identify the supercoordinates $(x^a, \theta^{\alpha}_i, \bar{\theta}^i_{\dot{\alpha}})$ with the standard quaternion $x$ and the Grassmann quaternions $\theta^+, \theta^-$. We also define the chiral and antichiral coordinates:
\begin{equation}
\label{eq:QuaternionicChiralCoord}
x^+ = x+\frac{1}{2}\bar{\theta}^- \theta^+, \quad x^- = \bar{x}+\frac{1}{2}\bar{\theta}^+ \theta^- \ ,
\end{equation}
which satisfy $x = (x^++\bar{x}^-)/2$. It is important to note that for a product of anticommuting quaternions $\alpha,\beta \in \mathbb{H}_a$ we get a minus sign under quaternion conjugation, i.e.  $\overline{\alpha \beta} = -\bar{\beta}\bar{\alpha}$.

The $\mathcal{N}=2$ superconformal transformations in Euclidean four dimensions can be written in terms of the quaternion supercoordinates $(x,\theta^+,\theta^-)$ in the following way:
\begin{itemize}
\item Supertranslations -  translations $P(a)$, $a\in \mathbb{H}$, and supercharges $Q_{\pm}\left(\xi^{\pm}\right)$, $\xi^\pm \in \mathbb{H}_a$:
\begin{equation}
\label{eq:QuatSupertrans}
x^+ \to x^+ + a + \bar{\xi}^-\theta^+, \quad x^- = x^- + \bar{a} + \bar{\xi}^+\theta^-, \quad \theta^+ \to \theta^+ + \xi^+, \quad \theta^- \to \theta^- + \xi^-  \ . 
\end{equation}
\item Dilatations $\lambda D$, $\lambda \in \mathbb{R}$: 
\begin{equation}
\label{eq:QuatDilat}
x^{\pm} \to \lambda x^{\pm},\quad \theta ^{\pm} \to \lambda ^{\frac{1}{2}} \theta^{\pm} \ .
\end{equation}
\item $\mathrm{O}(4)$ rotations $M(\omega^+,\omega^-)$, $\;\overline{\omega}^{\pm} = -\omega^{\pm} \in \mathbb{H}$: 
\begin{equation}
\label{eq:QuatO(4)}
x^+ \to e^{\omega^-}x^+ e^{-\omega^+}, \quad x^- \to e^{\omega^+}x^- e^{-\omega^-}, \quad \theta^+ \to \theta^+ e^{-\omega^+}, \quad \theta^{-} \to \theta^{-} e^{-\omega^-} \ .
\end{equation}
\item R-symmetry - $\mathrm{SU}(2)_R$ symmetry $G(q)$,  $\; \bar{q} = -q\in \mathbb{H}$, $\mathrm{O}(1,1)_R$ symmetry $\varphi A$, $\; \varphi \in \mathbb{R}$:
\begin{equation}
\label{eq:QuatRsymm}
x^{\pm} \to x^{\pm}, \quad \theta^+ \to e^{q+\frac{1}{2}\varphi}\theta^+, \quad \theta^- \to e^{q-\frac{1}{2}\varphi}\theta^- \ .
\end{equation}
\item Special conformal transformations $K(b)$, $b\in \mathbb{H}$:
\begin{equation}
\label{eq:QuatSuperconf}
\begin{aligned}
&x^+ \to x^+(1+\bar{b}x^+)^{-1}, \quad x^- \to x^+(1+bx^-)^{-1} \ , \\ &\theta^+ \to \theta^+(1+\bar{b}x^+)^{-1}, \quad \, \theta^- \to \theta^-(1+bx^-)^{-1} \ ,
\end{aligned}
\end{equation}
 and special superconformal supercharges $S_{\pm}\left(\eta^{\pm}\right)$ , $\eta^\pm \in \mathbb{H}_a$: 
 \begin{equation}
 \begin{aligned}
 &x^{\pm} \to x^{\pm}\left(1+\eta^{\pm}\theta^{\pm}\right)^{-1}, \quad x^{\mp} \to x^{\mp} , \\
 &\theta^{\pm} \to \theta^{\pm}\left(1+\eta^{\pm}\theta^{\pm}\right)^{-1}, \quad \theta^{\mp} \to \theta^{\mp} \mp \bar{\eta}^{\pm}x^{\mp}  \ .
 \end{aligned}
 \end{equation}
 
\end{itemize}
Note that in the $\mathcal{N}=2$ Euclidean superconformal group the abelian R-symmetry is given by the non-compact group $\mathrm{O}(1,1)_R$ instead of the compact $\mathrm{U}(1)_R$ present in the Lorentzian case. The generators of these transformations satisfy the required commutation relations listed in appendix \ref{app:QuaternionicAlgebra}. We define the superderivatives $D_{\pm}(\zeta^{\pm})$ by the requirements:
\begin{equation}
\begin{aligned}
[D_{+}(\zeta^+),D_{+}(\zeta'^+)] = [D_{-}(\zeta^-),D_{-}(\zeta'^-)] &= 0, \quad [D_{+}(\zeta^+),D_{-}(\zeta^-)] = P(\bar{\zeta}^-\zeta^+) \ , \\
[D_{\pm}(\zeta^{\pm}),Q_{\pm}(\xi^{\pm})] = [D_{\pm}(\zeta^{\pm}),Q_{\mp}(\xi^{\mp})] &=0 \ ,
\end{aligned}
\end{equation}
which are satisfied by the generators of the transformation:
\begin{equation}
x^{+} \to x^{+} + \bar{\theta}^- \zeta^+, \quad x^- \to x^- + \bar{\theta}^+\zeta^-, \quad \theta^+ \to \theta ^+ +\zeta^+, \quad \theta^- \to \theta ^- +\zeta^- \ .
\end{equation}
As a result, the chiral and antichiral coordinates satisfy the chirality conditions:
\begin{equation}
\label{eq:ChiralCoordCond}
D_- x^+ = D_+ x^- = 0 \ .
\end{equation}

A general superconformal transformations of the quaternion supercoordinates is a combination of the transformations \eqref{eq:QuatSupertrans}-\eqref{eq:QuatSuperconf} and therefore is given by a quaternionic super-M\"{o}bius transformation (quaternionic linear fractional transformations). In accordance with the isomorphism of the supercornformal group with $\mathrm{SL}(2|1;\mathbb{H})$, quaternionic super-M\"{o}bius transformations are parameterized by two supermatrices $M^+, M^- \in \mathrm{SL}(2|1;\mathbb{H})$ of the form \eqref{eq:SLquaternion}. The general quaternionic super-M\"{o}bius transformation is then given by:
\begin{equation}
\label{eq:SuperMobiusTrans}
\begin{aligned}
&x^{\pm} \to x'^{\pm}=\left(a^{\pm}x^{\pm}+b^{\pm}+\gamma^{\pm}\theta^{\pm}\right)\left(c^{\pm}x^{\pm}+d^{\pm}+\delta^{\pm}\theta^{\pm}\right)^{-1}  \ ,\\
&\theta^{\pm} \to \theta'^{\pm}=\left(\alpha^{\pm}x^{\pm}+\beta^{\pm}+e^{\pm}\theta^{\pm}\right)\left(c^{\pm}x^{\pm}+d^{\pm}+\delta^{\pm}\theta^{\pm}\right)^{-1}  \ .
\end{aligned}
\end{equation}
The two supermatrices $M^+, M^-$ are not independent, and one can be determined by the other. This is done by noting that the reality condition which follows from the definition of the chiral coordinates \eqref{eq:QuaternionicChiralCoord},
\begin{equation}
\label{eq:RealCond}
x'^+ - \bar{x}'^- = \bar{\theta}'^- \theta'^+ \ ,
\end{equation}
provides a relation between the supermatrices of the two chiralities.
The reality condition \eqref{eq:RealCond} results in equations relating the two sets of parameters:
\begin{equation}
\label{eq:RealCondSupermatrix}
\begin{pmatrix}
a^- & b^- & \gamma^- \\
c^- & d^- & \delta^- \\
\alpha^- & \beta^- & e^-
\end{pmatrix} = \begin{pmatrix}
\bar{d}^+ & -\bar{b}^+ & -\bar{\beta}^+ \\
-\bar{c}^+ & \bar{a}^+ & \bar{\alpha}^+ \\
\bar{\delta}^+ & -\bar{\gamma}^+ & \bar{e}^+
\end{pmatrix}^{-1} \ .
\end{equation}

 \subsection{Moduli Space in the Super-M\"{o}bius Formalism}

A chiral superfield $\Phi^+$ and an antichiral superfield $\Phi^-$ in the quaternionic formalism are defined by the chirality conditions:
\begin{equation}
\label{eq:QuatChiralSuperfieldCond}
D_- \Phi ^+(x^+,\theta^+) = 0, \quad D_+ \Phi ^- (x^-, \theta^-) = 0 \ ,
\end{equation}
where as a result of \eqref{eq:ChiralCoordCond} the superfields depend only on the corresponding chiral coordinates. We are interested in describing the moduli space of chiral superfield solutions to the field equations \eqref{eq:FieldEOM} using  the super-M\"{o}bius transformations. We first construct chiral superfield \eqref{eq:QuatChiralSuperfieldCond} corresponding to the solution \eqref{eq:SolToClassicalEOMRegularNotation} of the field equations. In quaternionic notation this solution takes the form:
\begin{equation}
\label{eq:QuaternChiralSuperfield}
\begin{aligned}
\Phi^+(x^+,\theta^+) &= A^+(x^+) +\bar{\theta}^+ B^+(x^+)\theta^++\theta^+\bar{\theta}^+\theta^+\bar{\theta}^+C^+(x^+) \ , \\
\Phi^-(x^-,\theta^-) &= A^-(x^-) +\bar{\theta}^- B^-(x^-)\theta^- +\theta^-\bar{\theta}^-\theta^-\bar{\theta}^-C^-(x^-)\ ,
\end{aligned}
\end{equation}
where $A^{\pm}(x^{\pm}),C^{\pm}(x^{\pm})\in \mathbb{R}$ are real fields and $\bar{B}^{\pm} (x^{\pm}) = - B^{\pm}(x^{\pm})\in \mathbb{H}$ are imaginary quaternionic fields, which ensures that the chiral and antichiral superfields are both real, i.e. $\bar{\Phi}^{\pm} = \Phi^{\pm}$. 

 The real degrees of freedom $A^+, A^-$ and  $C^+, C^-$ correspond to the complex fields $A$ and $C$ respectively, and the imaginary quaternionic degrees of freedom $B^+, B^-$ correspond to the complex  $\mathrm{SU}(2)_R$ triplet field $B_{ij}$. The fields $A^{\pm}, B^{\pm}, C^{\pm}$ are $\mathrm{O}(4)$ scalars, while under $\mathrm{SU}(2)_R$ symmetry \eqref{eq:QuatRsymm}, $A^{\pm},C^{\pm}$ are singlets and $B^{\pm}$ transform in the adjoint representation $B^{\pm} \to e^q B^{\pm} e^{-q}$. The super-Liouville chiral field transforms in the affine representation of the $O(1,1)_R$ symmetry, and therefore under the transformation \eqref{eq:QuatRsymm} we have $A^{\pm}(x^{\pm}) \to A^{\pm}(x^{\pm}) \pm \frac{1}{b} \varphi$,  $B^{\pm}(x^{\pm}) \to e^{\mp \varphi} B^{\pm}(x^{\pm})$ and $C^{\pm}(x^{\pm}) \to e^{\mp 2\varphi} C^{\pm}(x^{\pm})$.

The classical solution $\Phi_c=b\Phi$ to the field equations \eqref{eq:SolToClassicalEOMRegularNotation} can be written explicitly in terms of the quaternionic supercoordinates  as:
\begin{equation}
\label{eq:ABForQuat}
A_c^{\pm} = -\log\left( \frac{|x^{\pm}|^2+r^2}{2r^2}\right), \quad B_c^{\pm} = \frac{r B^{\pm}_0}{|x^{\pm}|^2+r^2} , \quad C_c^{\pm} = \frac{32r^2}{\left(|x^{\pm}|^2+r^2\right)^2} \ ,
\end{equation}
where $\overline{B}^{\pm}_0 = -B^{\pm}_0\in \mathbb{H}$ is a constant imaginary quaternion, which can be set to an arbitrary value using the $\mathrm{O}(1,1)_R \times \mathrm{SU}(2)_R$ symmetry.
The general solution to the field equations is given by the super-M\"{o}bius transformations of the basic superfield solution  \eqref{eq:ABForQuat}. The superconformal transformation introduces super-Weyl transformation, parameterized by the (anti)chiral superfield $\sigma^{\pm}= \log|c^{\pm}x^{\pm}+d^{\pm}+\delta^{\pm}\theta^{\pm}|^2$. Using the super-M\"{o}bius transformations \eqref{eq:SuperMobiusTrans} and the transformation law of the Liouville superfield under super-Weyl transformations \eqref{eq:PhiSWeylTrans} we find the general solution:
\begin{equation}
\label{eq:GenSolutionChiralSuperifledAfterTrans}
\begin{aligned}
\Phi_c'^{\pm} &= A^{\pm}_c(x'^{\pm})+\bar{\theta}'^{\pm}B_c^{\pm}(x'^{\pm})\theta'^{\pm}+\theta'^{\pm}\bar{\theta}'^{\pm}\theta'^{\pm}\bar{\theta}'^{\pm}C^{\pm}(x'^{\pm})-\sigma^{\pm}\\
&= \log\left(2r^2\right) -\log\left( |a^{\pm}x^{\pm}+b^{\pm}+\gamma^{\pm}\theta^{\pm}|^2+|c^{\pm}x^{\pm}+d^{\pm}+\delta^{\pm}\theta^{\pm}|^2\right)\\
& \quad +\frac{(\bar{x}^{\pm}\bar{\alpha}^{\pm}+\bar{\beta}^{\pm}+\bar{\theta}^{\pm}\bar{e}^{\pm})B_0^{\pm}(\alpha^{\pm}x^{\pm}+\beta^{\pm}+e^{\pm}\theta^{\pm})}{|a^{\pm}x^{\pm}+b^{\pm}+\gamma^{\pm}\theta^{\pm} |^2+|c^{\pm}x^{\pm}+d^{\pm}+\delta^{\pm}\theta^{\pm}|^2} \\
& \quad + \frac{32\left|\alpha^{\pm}x^{\pm}+\beta^{\pm}+e^{\pm}\theta^{\pm}\right|^4}{\left(|a^{\pm}x^{\pm}+b^{\pm}+\gamma^{\pm}\theta^{\pm} |^2+|c^{\pm}x^{\pm}+d^{\pm}+\delta^{\pm}\theta^{\pm}|^2\right)^2}\ .
\end{aligned}
\end{equation}

\subsection{Correlation Functions of Light  Operators}

In the semiclassical limit $b\to 0$, the correlation function of light operators is evaluated using the saddle point approximation \eqref{eq:SemiclassCorrExp} by integrating over the moduli space of solutions to the field equations \eqref{eq:FieldEOM}. Looking at the general solution written using the quaternionic formalism
\eqref{eq:GenSolutionChiralSuperifledAfterTrans} we see that the moduli space is $\mathcal{M}=\mathrm{SL}(2|1;\mathbb{H})$. In order to get the leading exponential asymptotic of the correlation function of light operators we need to include two corrections to \eqref{eq:SemiclassCorrExp}. We need to to multiply by the functional superdeterminant $\sdet\left(\frac{\delta^2 S(\Phi_c)}{\delta \Phi^2}\right)$, and we also need to multiply by the superdeterminant of the Jacobian for changing the integration variable $\Phi_c$ to the coordinates of the moduli space. We can include both those effects by multiplying \eqref{eq:GenSolutionChiralSuperifledAfterTrans} by a $b$-dependent factor $\hat{\mathcal{A}}(b)$ whose logarithm is at most $O(\log b)$
\cite{Harlow:2011ny}. We therefore have the following semiclassical expression for the correlation function of light operators:
\begin{equation}
\label{eq:SemiclassCorrExpMobius}
\left<\mathcal{V}_{b\sigma_1 b\tilde{\sigma}_1}(z_1)\cdots \mathcal{V}_{b\sigma_N b\tilde{\sigma}_N}(z_N)\right> \approx  \hat{\mathcal{A}}(b) e^{-S_L(\Phi_c,\Phi_c)} \int_{\mathcal{M}} d\mu(M)\,\prod_{l=1}^{N} e^{2\sigma_l\Phi^+_M(z_l)+2\tilde{\sigma}_l\Phi^-_{M}(z_l)} \ ,
\end{equation}
where $\Phi^{\pm}_M$ denotes the saddle point $\eqref{eq:GenSolutionChiralSuperifledAfterTrans}$ corresponding to $M^{\pm}$ and the integration is accompanied by the invariant Haar measure of quaternionic supergroup $\mathrm{SL}(2,1;\mathbb{H})$ \eqref{eq:SLquaternion}:
\begin{equation}
\label{eq:SLHaarMeasure}
d\mu(M^+) = d^4a^+\, d^4b^+\, d^4c^+\, d^4d^+\, d^4\alpha^+\, d^4\beta^+\, d^4\gamma^+\, d^4\delta^+\, d^4e^+\; \delta\!\left(\sdet\left(M^+\right)-1\right) \ .
\end{equation}
Here we wrote the Haar measure in terms of elements of the supermatrix $M^+$ but it takes the same form in terms of $M^-$.

In order for a correlation function to be non-zero, it needs to have a well-defined R-charge (see e.g. \cite{Osborn:1998qu} for the $\mathcal{N}=1$ case). The $\mathrm{O}(1,1)_R$ charge of the vertex operator \eqref{eq:GeneralVertexInSuperspace} is the Euclidean version of the Lorentzian non-compact $\mathrm{U}(1)_R$ charge \eqref{eq:U1RGenVertexInSuperspace}, and is given by the same formula. The sum over all $\mathrm{O}(1,1)_R$-charges of the vertex operators which appear in a given correlation function of the form \eqref{eq:DefCorrMain} should agree with a series expansion in superspace coordinates which contains only integer powers of the Grassmanic external coordinates. Thus the $N$-point function of vertex operators \eqref{eq:GeneralVertexInSuperspace} vanishes unless it obeys the $\mathrm{O}(1,1)_R$ selection rule:
\begin{equation}
\frac{2}{b}\sum_{l=1}^{N} (\alpha_l - \tilde{\alpha}_l) \in \mathbb{Z} \ .
\end{equation}

Our strategy is to start with the classical solution for the chiral superfield given by \eqref{eq:QuaternChiralSuperfield}-\eqref{eq:ABForQuat} and conformally transform it according to \eqref{eq:GenSolutionChiralSuperifledAfterTrans}. Then, one must integrate over the parameters of the $\mathcal{N}=2$ superconformal transformation to find the semiclassical limit of the correlation function of vertex operators.

We are interested in understanding how  the $\mathrm{U}(1)_R$ selection rule arises from the integration over the $\mathrm{SL}(2,1;\mathbb{H})$ moduli space of saddle points. In terms of the supermatrix $M^+$, an $\mathrm{O}(1,1)_R$ transformation is given by multiplying $M^+$ by an $\mathrm{SL}(2,1;\mathbb{H})$ element
\begin{equation}
\label{eq:SupermatrixO(1,1)R}
M^+ \to  M^+ \begin{pmatrix}
e^{\varphi} & 0 & 0 \\
0 & e^{\varphi} & 0 \\
0 & 0 & e^{2\varphi}
\end{pmatrix} \ ,
\end{equation}
which, of course, leaves the Haar measure \eqref{eq:SLHaarMeasure} invariant. On the other hand, for the vertex operators evaluated at the solution \eqref{eq:GenSolutionChiralSuperifledAfterTrans} we have :
\begin{equation}
\exp\left(2\sigma\Phi^++ 2\tilde{\sigma} \Phi^-\right) \to e^{-2(\sigma-\tilde{\sigma})\varphi} \exp\left(2\sigma\Phi^+(x^+,e^{\varphi}\theta^+)+ 2\tilde{\sigma} \Phi^-(x^-,e^{-\varphi}\theta^-)\right) \ .
\end{equation}
By expanding the vertex operators appearing in a semiclassical calculation of an $N$-point function in  powers of the $\theta^{\pm}_l, \; l=1,\dots, N$, we find that under \eqref{eq:SupermatrixO(1,1)R} each order of the power series is multiplied by $\exp\left(-2\sum_l (\sigma_l - \tilde{\sigma}_l)\varphi+k\varphi\right)$ for some $k\in \mathbb{Z}$. However, \eqref{eq:SupermatrixO(1,1)R} is simply a change of integration variables and thus cannot change the result of the integral. Therefore, we have shown that the result of the integration over the moduli space will have a defined $\mathrm{O}(1,1)_R$ charge and demonstrated the selection rule $2\sum_l (\sigma_l - \tilde{\sigma}_l) \in \mathbb{Z}$.

In addition to the reproduction of the $\mathrm{O}(1,1)_R$ selection rule, by explicitly inserting the classical solution \eqref{eq:GenSolutionChiralSuperifledAfterTrans} in the moduli space integral 
\eqref{eq:SemiclassCorrExpMobius}, we find some non-trivial selection rules which exist in the semiclassical limit. Specifically, by examining the lowest order component in all supercoordinates of \eqref{eq:SemiclassCorrExpMobius}, i.e. $\theta_1=\bar{\theta}_1 = \dots = \theta_N = \bar{\theta}_N=0$, one can see that the 2-point and 3-point functions of the chiral and antichiral vertex operators $V_{\alpha} = e^{2\alpha A}, V_{\tilde{\alpha}} = e^{2\tilde{\alpha} \bar{A}}$ vanish in the semiclassical computation. This vanishing is a result of the number of Grassmann integrations appearing in the measure \eqref{eq:SLHaarMeasure}. By preforming a series of changes to the integration variables for these 2-point and 3-point functions one can show that there is not a sufficient number of Grassmann variables appearing in the integrand for it to survive the integrations.

\section{Discussion and Outlook}
\label{sec:SummaryAndOutlook}

We constructed and studied  classical and quantum aspects of $\mathcal{N}=2$ Liouville SCFT in four dimensions.
There are many directions one can follow. 
As in the $\mathcal{N}=1$ case, solving analytically the integrals for the three-point function of vertex operators should reveal a four-dimensional DOZZ-like formula, which
in turn can lead to a complete bootstrap solution of the theory.  

Solving the integrals over the quaternionic variables, which we largely left as an open problem, will result
in the explicit expression of correlation functions of the vertex operators in the semiclassical regime. 
In fact, the quaternionic formalism presented in subsection \ref{subsec:ForDNequals2TransQuanternions} for general $\mathcal{N}=2$ superconformal transformations, which is based on the isomorphism of the supercornformal group with $\mathrm{SL}(2|1;\mathbb{H})$ that was proven in \cite{Lukierski:1983jg}, can be used for other calculations in a general framework of $\mathcal{N}=2$ SCFTs, as it leads to a simple, quaternionic linear fractional form of transformations. In \cite{Levy:2018xpu}, it was shown that by using the super-M\"{o}bius group one can easily evaluate correlation functions of vertex operators in the semiclassical limit. In this work, a quaternionic super-M\"{o}bius group was shown to yield some light over these correlators, however, the complete calculation had faced some technical difficulties. It will be interesting to develop new mathematical frameworks  and other generalizations of super-M\"{o}bius groups for evaluating correlation functions of vertex operators in the semiclassical limit in other Liouville field theories in various dimensions.

In \cite{Beem:2013sza}, the authors have established a correspondence between four-dimensional $\mathcal{N}=2$ SCFTs and two-dimensional chiral algebras. By classifying the Schur operators in the four dimensional theory, it was argued that in cases where the four-dimensional $\mathcal{N}=2$ theory is unitary, a component of the $\mathrm{SU}(2)_R$ current yields the stress-tensor of the corresponding two-dimensional theory. A Schur operator \cite{Gadde:2011uv} is an operator satisfying:
\begin{equation}
\label{eq:SchurCon1}
\frac{1}{2}\left(\Delta-(j_1+j_2)\right)-R=0, 
\end{equation}
\begin{equation}
\label{eq:SchurCon2}
w+(j_1-j_2)=0.
\end{equation}
It was shown in \cite{Beem:2013sza} that when the four-dimensional theory is unitary, the second condition necessarily follows from the first one. However, in non-unitary $\mathcal{N}=2$ four-dimensional SCFTs, such as the theory that was studied in this paper, the two conditions are independent. Due to the non-unitarity of the theory it is possible to find operators which are Schur operators but still transform trivially under the $\mathrm{SU}(2)_R$ symmetry (see table \ref{Table1} for a classification of the dimensions and charges of the various fields). For example, $\partial_{++}A$, and $F_{++}$\footnote{We denote by $F_{++}$ and $\partial_{++}A$ the operators which are highest weight states of $\mathrm{SU}(2)_1$  (see table \ref{Table1}).  } are two operators which satisfy the Schur conditions \eqref{eq:SchurCon1} and \eqref{eq:SchurCon2}. When reducing their two-point functions in flat space according to the prescription given in \cite{Beem:2013sza}, one finds non-vanishing free field correlators in two dimensions. However, both of these operators transform trivially under the $\mathrm{SU}(2)_R$ group, therefore do not appear in the $\mathrm{SU}(2)_R$ current. It will be interesting to study how does the loss of unitarity in the four-dimensional theory affects the correspondence with the two-dimensional chiral algebra.

\vskip 1cm
{\bf \large Acknowledgement } 
\vskip 0.5cm
This work is supported in part by the I-CORE program
of Planning and Budgeting Committee (grant number 1937/12), the US-Israel Binational Science Foundation, GIF and the ISF Center of Excellence. T.L gratefully acknowledges the support of the Alexander Zaks Scholarship. A.R.M gratefully acknowledges the support of the Adams Fellowship Program of the Israel Academy of Sciences and Humanities.

\appendix

\section{Notations and Conventions}
\label{app:Notations}
\subsection{General Notations and Conventions}
The flat space notations and component reduction are inherited mainly from \cite{Butter:2013lta} in Lorentzian signature. These were adapted to Euclidean signature using appendix A in \cite{Festuccia:2011ws}. The curved supergravity notation follows that of \cite{Kuzenko:2013gva} and \cite{Kuzenko:2008ep} in Lorentzian signature.

Spacetime indices are denoted by $a,b,\cdots$, while $SU(2)_R$ indices are denoted by $i,j,\cdots$ and
spinoric indices are denoted by $\alpha,\beta,\cdots$. \\
$SU(2)_R$ indices are raised and lowered by complex conjugation, in accordance with \cite{Butter:2013lta}. The invariant $SU(2)_R$ tensor $\epsilon^{ij}$ and $\epsilon_{ij}$ are defined by $\epsilon^{ij}\epsilon_{kj}=\delta^i_k$, with $\epsilon^{12}=\epsilon_{12}=1$.

Spinoric indices $\alpha,\dot{\alpha}=\lbrace1,2\rbrace$ are raised and lowered using the antisymmetric $\epsilon$ symbol:
\begin{equation}
\psi^\alpha=\epsilon^{\alpha\beta}\psi_\beta, \quad \psi_\alpha=\epsilon_{\alpha\beta}\psi^\beta, \quad \bar{\psi}^{\dot{\alpha}}=\epsilon^{\dot{\alpha}\dot{\beta}}\bar{\psi}_{\dot{\beta}}, \quad \bar{\psi}_{\dot{\alpha}}=\epsilon_{\dot{\alpha}\dot{\beta}}\psi^{\dot{\beta}},
\end{equation}
where $\epsilon^{12}=\epsilon_{21}=1$. 
A four component Dirac fermion $\Psi$ consists of two Weyl spinors $\psi_\alpha$ and $\bar{\chi}^{\dot{\alpha}}$ which are left-handed and right handed spinors respectively. The Dirac conjugate is $\bar{\Psi}$ follows the notation in \cite{Butter:2013lta} and carries components $\chi^\alpha=(\bar{\chi}^{\dot{\alpha}})^*$ and $\bar{\psi}_{\dot{\alpha}}=(\psi_\alpha)^*$.  

We denote:
\begin{equation}
V_{\alpha\dot{\alpha}} = (\sigma^a)_{\alpha\dot{\alpha}}V_{a}, \qquad V_a=-2\bar{\sigma}_a^{\dot{\alpha}\alpha}V_{\alpha\dot{\alpha}},
\end{equation}
and:
\begin{equation}
F^-_{ab}=(\sigma^{ab})_{\alpha}^\beta F_\beta^\alpha,\qquad F^+_{ab}=(\bar{\sigma}_{ab})^{\dot{\alpha}}_{\dot{\beta}}F^{\dot{\beta}}_{\dot{\alpha}},
\end{equation}
\begin{equation}
F^{\pm}_{ab} = \frac{1}{2}\left(F_{ab}\pm\tilde{F}_{ab} \right), \quad \tilde{F}_{ab}=\frac{1}{2}\epsilon_{abcd}F^{cd}, \quad \tilde{F}^{\pm}_{ab}=\pm F^{\pm}_{ab}.
\end{equation}

Supercoordinates are denoted by $z^A=(x^a,\theta^{\alpha i},\bar{\theta}_{\dot{\alpha}i})$. 
In Lorentzian signature, the superderivatives in flat space are given by:
\begin{equation}
D_{\alpha i} = \frac{\partial}{\partial\theta^{\alpha i}}+i(\sigma^a)_{\alpha\dot{\alpha}}\bar{\theta}^{\dot{\alpha}}_i\frac{\partial}{\partial x^a}, \qquad \bar{D}^{\dot{\alpha}i} = \frac{\partial}{\partial\bar{\theta}_{\dot{\alpha}i}}+i(\bar{\sigma}^a)^{\alpha\dot{\alpha}}\theta_\alpha^i\frac{\partial}{\partial x^a}.
\end{equation}
An $\mathcal{N}=2$ chiral superfield satisfies:
\begin{equation}
\bar{D}^{\dot{\alpha}i}\Phi=0.
\end{equation}
In terms of component fields, it decomposes into:
\begin{equation}
\begin{aligned}
&\qquad A \equiv \Phi\vert_{\theta=0}, \qquad \Psi_{\alpha i} \equiv D_{\alpha i} \Phi\vert_{\theta=0}, \qquad B_{ij}\equiv -\frac{1}{2}D_{ij}\Phi\vert_{\theta=0},\\
& F_{ab}^-\equiv -\frac{1}{4}(\sigma_{ab})_\alpha^\beta D_\beta^\alpha\Phi\vert_{\theta=0},\qquad \Lambda_{\alpha i} \equiv \frac{1}{6}\epsilon^{jk}D_{\alpha k}D_{ji}\Phi\vert_{\theta=0},\qquad C=-2D^4\Phi\vert_{\theta=0}, 
\end{aligned}
\end{equation}
where 
\begin{equation}
D_{ij}\equiv -D_{\alpha (i}D^\alpha_{j)}, \qquad D_{\alpha\beta}\equiv -\epsilon^{ij}D_{(\alpha i}D_{\beta)j}.
\end{equation}
Note that in \eqref{eq:KineticComponents} we have used the normalization found in \cite{deWit:1980lyi} for the kinetic term, which is different than the one given in \cite{Butter:2013lta} by a factor of $4$.

\subsection{Notation for Grassmann Quaternions and Useful Identities}
\label{app:QuaternionicNotation}
Consider a Grassmanic quaternion $\alpha$. We denote its components by $\alpha_0 {\bf 1}+\alpha_1{\bf i}+\alpha_2{\bf j}+\alpha_3{\bf k}$, where $\alpha_0,\alpha_1,\alpha_2,\alpha_3$ are real Grassman numbers. We define:
\begin{equation}
|\alpha|^2 \equiv \alpha\bar{\alpha}.
\end{equation}
Therefore:
\begin{equation}
|\alpha|^4 = -24\alpha_0\alpha_1\alpha_2\alpha_3.
\end{equation} 
Note that the following identity holds for Grassmann quaternions:
\begin{equation}
|\beta|^2|(\alpha+\beta)|^4|\alpha|^2 = |\alpha|^4|\beta|^4.
\end{equation}
The latter is easily shown by defining $\gamma=\alpha+\beta$: 
\begin{equation}
|\beta|^2|\alpha+\beta|^4|\alpha|^2  = |\gamma-\alpha|^2|\gamma|^4|\alpha|^2=|\alpha|^2|\gamma|^4|\alpha|^2 = |\alpha|^2|\beta|^4|\alpha|^2=|\alpha|^4|\beta|^4
\end{equation}
where we have used the fact that for any Grassmanic quaternion $\alpha$: $|\alpha|^n=0$ for $n>4$.

\section{Superspace Propagators}
\label{app:SuperSpacePropagators}

In the presence of chiral sources $J$, $\bar{J}$ the free action can be written as:
\begin{equation}
\begin{aligned}
S &= \frac{1}{32\pi^2}\int d^4xd^4\theta d^4\bar{\theta}\Phi\bar{\Phi}+\left(\int d^4xd^4\theta\Phi J +h.c.\right)\\
&=\frac{1}{32\pi^2} \frac{1}{2}\begin{pmatrix}
\Phi, & \bar{\Phi}
\end{pmatrix}\begin{pmatrix} 
0 & 1 \\
1 & 0 
\end{pmatrix} \begin{pmatrix}
\Phi \\ 
\bar{\Phi}
\end{pmatrix} + \begin{pmatrix}
\Phi, & \bar{\Phi}
\end{pmatrix}\begin{pmatrix} 
\frac{D^4}{\square^2} & 0 \\
0 & \frac{\bar{D}^4}{\square^2} 
\end{pmatrix} \begin{pmatrix}
J \\ 
\bar{J}
\end{pmatrix}
\end{aligned}
\end{equation}
where we have used $\int d^4 \bar{\theta} = \bar{D}^4$, and the identity:
\begin{equation}
\frac{\bar{D}^4 D^4}{\square^2}\Phi = \Phi ,
\end{equation}
which holds for any $\mathcal{N}=2$ chiral superfield $\Phi$. The variation is given by:
\begin{equation}
\frac{\delta\Phi(y,\theta^i)}{\delta\Phi(y',\theta'_i)}= \delta(y-y')\delta(\theta^i-\theta'^i), \qquad \frac{\delta\Phi(x,\theta^i,\bar{\theta}_i)}{\delta\Phi(x',\theta'^i,\bar{\theta}'_i}=\bar{D}^4\delta(z-z'),
\end{equation}
where 
\begin{equation}
\delta(z-z') \equiv \delta(x-x')\delta(\theta^i-\theta'^i)\delta(\bar{\theta}_i-\bar{\theta}'_i).
\end{equation}
We thus find:
\begin{equation}
\frac{1}{32\pi^2}\begin{pmatrix}
\bar{D}^4 & 0 \\
0 & D^4
\end{pmatrix}\begin{pmatrix}
0 & 1\\
1 & 0
\end{pmatrix} \begin{pmatrix}
\Phi \\
\bar{\Phi}\end{pmatrix} = - \begin{pmatrix}
J \\
\bar{J}
\end{pmatrix}.
\end{equation}
The Green's function satisfies:
\begin{equation}
\frac{1}{32\pi^2}\begin{pmatrix}
\bar{D}^4 & 0 \\
0 & D^4
\end{pmatrix} \begin{pmatrix}
0 & 1 \\
1 & 0 
\end{pmatrix} \vartriangle = \begin{pmatrix}
\bar{D}^4 & 0 \\
0 & D^4 
\end{pmatrix} \delta(z-z').
\end{equation}
A solution to this equation is given by:
\begin{equation}
\vartriangle = 32\pi^2 \begin{pmatrix}
0 & \frac{\bar{D}^4D^4}{\square^2}\\
\frac{D^4\bar{D}^4}{\square^2} & 0 
\end{pmatrix}\delta(z-z'),
\end{equation}
therefore:
\begin{equation}
\vartriangle \equiv \begin{pmatrix}
\left<\Phi(z)\Phi(z')\right> & \left<\Phi(z)\bar{\Phi}(z')\right>\\
\left<\bar{\Phi}(z)\Phi(z')\right> & \left<\bar{\Phi}(z)\bar{\Phi}(z')\right> \end{pmatrix}= \begin{pmatrix}
0 & \bar{D}^4D^4 \\
D^4\bar{D}^4 & 0 
\end{pmatrix}\hat{\vartriangle}(x-x'),
\end{equation}
where $\hat{\vartriangle}(x-x')\equiv -\log{(x-x')}$.
Therefore one recovers the results given in equations \eqref{eq:PropSuper01}, \eqref{eq:PropSuper02}, \eqref{eq:PropSuperfields}.

\section{Quaternionic Formalism for $\mathcal{N}=2$ Superconformal Algebra in $4d$}
\label{app:QuaternionicAlgebra}

By acting with the $\mathcal{N}=2$ transformations on each of the quaternionic  supercoordinates (see subsection \ref{subsec:ForDNequals2TransQuanternions}), one recovers the algebra of \cite{Lukierski:1983jg}:

\begin{align}
& \left[ D,Q_2(\zeta) \right] = -\frac{1}{2}Q_2(\zeta), \qquad \left[D,\pi_1(\zeta)\right] = -\frac{1}{2}\pi_1(\zeta), \\
& \left[ D, Q_1(\zeta)\right]=\frac{1}{2}Q_z(\zeta), \qquad \left[D,\pi_2(\zeta)\right]=\frac{1}{2}\pi_2(\zeta), \\
& \left[ M(\omega_1, \omega_2), Q_2(\zeta)\right] =Q_2(\omega_2\zeta), \qquad \left[M(\omega_1,\omega_2),\pi_1(\zeta)\right] = \pi_1(\omega_1,\zeta), \\
& \left[ M(\omega_1,\omega_2), Q_1(\zeta)\right] = Q_1(\omega_1\zeta), \qquad \left[M(\omega_1,\omega_2),\pi_2(\zeta)\right] = \pi_2(\omega_2\zeta), \\
& \left[ K(b),Q_2(\zeta)\right] = Q_1(\bar{b}\zeta), \qquad \left[ K(b),\pi_1(\zeta)\right] = -\pi_2(b\zeta), \\
& \left[ P(a),\pi_2(\zeta) \right] = -\pi_1(\bar{a}\zeta), \qquad \left[P(a),Q_1(\zeta) \right] = -Q_2(a\zeta),\\
& \left[ M(\omega_1,\omega_2), M(\omega_1',\omega_2')\right] = M(\left[\omega_1,\omega_1')\right],\left[\omega_2,\omega_2'\right]),\\
& \left[M(\omega_1,\omega_2),P(a)\right] = P(\omega_2a-a\omega_1), \\
& \left[ P(a),D\right] = P(a), \\
&\left[K(b),D\right] = -K(b), \\
&\left[ M(\omega_1,\omega_2),K(b)\right] = K(\omega_2b-b\omega_1), \\
& \left[ P(a),K(b)\right] = -(\bar{a}b+\bar{b}a)D+\frac{1}{2}M(\bar{a}b-\bar{b}a,a\bar{b}-b\bar{a}),
\end{align}
\begin{align}
&\left[G(q),Q_1(\zeta)\right]=-Q_1(\zeta q), \qquad \left[G(q),Q_2(\zeta) \right] = -Q_2(\zeta q), \\
& \left[ G(q), \pi_1(\eta) \right] = -\pi_1(\eta q), \qquad \left[ G(q), \pi_2(\eta) \right] = -\pi_2(\eta q),
\end{align}
\begin{align}
& \left[ A, Q_1(\zeta) \right] = -Q_1(\zeta), \qquad \left[ A, Q_2(\zeta) \right]=-Q_2(\zeta),\\
& \left[ A, \pi_1(\eta)\right] = \pi_1(\eta), \qquad \left[ A, \pi_2(\eta)\right] = \pi_2(\eta),
\end{align}

\begin{align}
& \lbrace Q_2(\zeta), \pi_1(\eta) \rbrace = P(\zeta\bar{\eta}), \\
& \lbrace Q_1(\zeta), \pi_2(\eta) \rbrace = K(\eta\bar{\zeta}), \\ 
& \lbrace Q_1(\zeta),\pi_1(\eta) \rbrace = -\frac{1}{2}G(\bar{\eta}\zeta-\bar{\zeta}\eta)-\frac{3}{4}\left(\zeta\bar{\eta}+\eta\bar{\zeta}\right)A\\
& \qquad \qquad \qquad  \qquad +\frac{1}{2}\left(\zeta\bar{\eta}+\eta\bar{\zeta}\right)D+\frac{1}{2}M(\zeta\bar{\eta}-\eta\bar{\zeta},0),\nonumber \\
& \lbrace Q_2(\zeta), \pi_2(\eta) \rbrace = -\frac{1}{2}G(\bar{\eta}\zeta-\bar{\zeta}\eta)-\frac{3}{4}\left(\bar{\eta}\zeta+\bar{\zeta}\eta\right)A\\
& \qquad \qquad \qquad \qquad -\frac{1}{2}\left(\bar{\eta}\zeta+\bar{\zeta}\eta\right)D+\frac{1}{2}M(0,\zeta\bar{\eta}-\eta\bar{\zeta}).\nonumber 
\end{align}

\section{Dimensions and Charges}

In this appendix we list the dimensions and charges of the fields in \eqref{eq:KineticComponents}. $E$ represents the dimensions, $j_1$ and $j_2$ are the corresponding  $\mathrm{O}(4)\cong \mathrm{SU}(2)_1\times\mathrm{SU}(2)_2$ charges, $R$ is the $\mathrm{SU}(2)_R$ charge and $r$ corresponds to the $\mathrm{U}(1)_R$ charge. The list is given in table \ref{Table1}.

\begin{table}[!ht]
\begin{center}
 \begin{tabular}{|c c c c c c |} 
 \hline
  & $\Delta$ & $j_1$ & $j_2$ & $R$ & $w$ \\ [0.5ex] 
 \hline\hline
 $F_{\alpha\beta}$ &$1$& $\pm 1,0$ & $0$ & $0$ & $-1$ \\ 
 \hline
$\bar{F}_{\dot{\alpha}\dot{\beta}}$& $1$&  $0$ & $\pm 1,0$ & $0$ & $1$ \\
 \hline
$B^{ij}$ & $1$& $0$ & $0$ & $\pm 1,0$ & $-1$ \\
 \hline
 $\bar{B}_{ij}$ & $1$& $0$ & $0$ & $\pm 1,0$ & $1$ \\
 \hline
$\psi^i_{\alpha}$ & $\frac{1}{2}$ & $\pm\frac{1}{2}$ & $0$ & $\pm \frac{1}{2}$ & $-\frac{1}{2}$ \\
\hline
$\bar{\psi}_i^{\dot{\alpha}}$ & $\frac{1}{2}$ & $0$ & $\pm \frac{1}{2}$ & $\pm\frac{1}{2}$ & $\frac{1}{2}$ \\
\hline
$\Lambda^i_{\alpha}$ & $\frac{3}{2}$& $\pm \frac{1}{2}$ & $0$ & $\pm\frac{1}{2}$ & $-\frac{3}{2}$ \\
\hline
$\bar{\Lambda}_i^{\dot{\alpha}}$ & $\frac{3}{2}$& $0 $ & $\pm\frac{1}{2}$ & $\pm\frac{1}{2}$ & $\frac{3}{2}$ \\
\hline
$e^{2\alpha A+2\tilde{\alpha}{A^*}}$ & $-4\alpha\tilde{\alpha}+\frac{2}{b}(\alpha+\tilde{\alpha})$& $0$ & $0$ & $0$ & $\frac{2}{b}(\alpha-\tilde{\alpha})$ \\
\hline
$ \partial_{\alpha\dot{\alpha}}A\, (Q=0)$ & $1$& $\frac{1}{2}$ & $\frac{1}{2}$ & $0$ & $0$ \\
\hline
$ \partial^{\alpha\dot{\alpha}}\bar{A}\, (Q=0)$ & $1$& $\frac{1}{2}$ & $\frac{1}{2}$ & $0$ & $0$ \\
[1ex] 
 \hline
\end{tabular} 
\caption{The dimensions and charges of the fields which correspond to the action in \eqref{eq:KineticComponents}. The last two lines represent primary operators only for the Coulomb gas case ($Q=0)$.} \label{Table1}
\end{center}
\end{table}

\end{document}